\documentclass[conference]{IEEEtran}
\IEEEoverridecommandlockouts
\usepackage{cite}
\usepackage{amsmath,amssymb,amsfonts}
\usepackage{algorithmic}
\usepackage{graphicx}
\usepackage{textcomp}
\usepackage[dvipsnames]{xcolor}
\usepackage{etoolbox}
\usepackage{booktabs}
\usepackage{url}
\usepackage{tabularx}
\usepackage{balance}
\usepackage{listings}

\def\BibTeX{{\rm B\kern-.05em{\sc i\kern-.025em b}\kern-.08em
    T\kern-.1667em\lower.7ex\hbox{E}\kern-.125emX}}
\begin{document}

% \title{Audio and Speech Coding Quality Benchmark
\title{OpenACE: An Open Benchmark for Evaluating Audio Coding Performance}

\author{\IEEEauthorblockN{Jozef Coldenhoff}
\IEEEauthorblockA{
\textit{Logitech Europe S.A.}\\
Lausanne, Switzerland}
\and
\IEEEauthorblockN{Niclas Granqvist}
\IEEEauthorblockA{
\textit{Logitech Europe S.A.}\\
Lausanne, Switzerland}
\and
\IEEEauthorblockN{Milos Cernak}
\IEEEauthorblockA{
\textit{Logitech Europe S.A.}\\
Lausanne, Switzerland
% \\ milos.cernak@ieee.org
}
% \and
% \and
% \and
}

\maketitle

\begin{abstract}
Audio and speech coding lack unified evaluation and open-source testing. Many candidate systems were evaluated on proprietary, non-reproducible, or small data, and machine learning-based codecs are often tested on datasets with similar distributions as trained on, which is unfairly compared to digital signal processing-based codecs that usually work well with unseen data. This paper presents a full-band audio and speech coding quality benchmark with more variable content types, including traditional open test vectors. An example use case of audio coding quality assessment is presented with open-source Opus, 3GPP’s EVS, and recent ETSI’s LC3 with LC3+ used in Bluetooth LE Audio profiles. Besides, quality variations of emotional speech encoding at 16 kbps are shown. The proposed open-source benchmark contributes to audio and speech coding democratization and is available at \url{https://github.com/JozefColdenhoff/OpenACE}.
\end{abstract}


\section{Introduction}
\label{sec:intro}
% The story of the paper:
% \begin{itemize}
%     \item Introduce DSP-based standardized EVS, LC3, LC3+, and open, royalty-free, Opus  codecs.
%     \item For completeness, introduce new trends with data-driven neural and hybrid codecs and challenges with their assessment.
%     \item Addressing the assessment challenge: assess the codecs on more realistic (real-life) stimuli, not just clean speech.
%     \item The goal of this paper is two-fold:: First, assess LC3 with EVS and Opus on the open benchmark, 48kHz clean speech, using POLQA, and 3QUEST: various bit-rates vs. quality.
%     \item Second, assess LC3, EVS and Opus subjectively using MUSRHA, on distorted speech, music, and communication channel (packet loss).
%     \item look at https://listening-test.coresv.net/results.htm
%     \item look at https://hydrogenaud.io/index.php/topic,122575.0.html
% \end{itemize}

Audio and speech coding is a key component of communication and streaming systems used ubiquitously worldwide. Model-based compression methods developed decades ago and continuously improved, extensively updated, and tested by telecom companies and certification institutions have faced a paradigm shift in the last decade due to data-driven approaches fueled by machine learning (ML). Traditional DSP (Digital Signal Processing) and neural codecs are now combined into hybrid systems~\cite{kim2024neural} and are often assessed together. %In industry, classic standardized and certified intrusive audio assessments prevail, but the ML-based non-intrusive speech quality assessment is almost the norm today. In both cases, subjective testing still provides the only reliable evaluation metrics.

Audio and speech coding is democratized by recent competitive neural codec development, released by various small research and startup teams, using available open-source or popular licensed data. There are several challenges in properly evaluating new neural audio codecs: (i) valid comparison to standardized codecs that are extensively tested but usually with proprietary test sequences, (ii) valid comparison between the new codecs themselves coming from different teams, and (iii) comparisons focusing either on achieved audio quality, delay or computational efficiency. Table~\ref{tab:codecs} outlines the databases used to evaluate recent neural codecs. First, almost every codec was evaluated using different data. Second, some datasets have restricted access; third, most contain only English language sequences. Often, the neural codecs are not tested cross-datasets, which is not fair compared to the DSP-based codecs, which are not optimized for particular data. 

% \footnotesize
\begin{table}[t!]
\caption{Neural codecs and their evaluation speech data. SoundStream uses also Freesound \cite{fonseca2017freesound} noises and music from MagnaTagATune~\cite{law2009evaluation}. DAPS evaluates also on music from MUSDB dataset~\cite{MUSDB18}  N/O: Non-Open.}
\centering
\begin{tabular}{ccccccccc}
\toprule
Codec & Dataset & Access & Language\\
\midrule
DAC~\cite{kumar2024high} & AudioSet~\cite{gemmeke2017audio} & Open & English \\
& DAPS~\cite{mysore2014can} & Open & English \\
SoundStream~\cite{zeghidour2021soundstream} & VCTK \cite{valentini2017noisy} & Open & English\\
& \small LibriTTS \cite{zen2019libritts} & Open & English\\
Lyra \cite{kleijn2021generative} & VCTK & Open & English\\
Zhen et al. \cite{zhen2021scalable} & TIMIT \cite{garofolo1993timit} & N/O & English\\
WaveNet \cite{kleijn2018wavenet} & WSJ []& N/O & English\\
LPCNet \cite{valin2019lpcnet} & NTT \cite{ntt1994cdrom} & N/O & Various\\
Gârbacea et al. \cite{garbacea2019WaveNet} &  LS \cite{panayotov2015LS} & Open & English\\
Kankanahalli \cite{kankanahalli2018} & TIMIT & N/O & English \\
\bottomrule
\end{tabular}
\label{tab:codecs}
\end{table}
\normalsize

Early attempts (2010--2016) of ML-based speech coding focused on binary quantization of speech representations, e.g.~\cite{deng2010binary,cernak2015phonological,cernak2016composition}, have been followed by recent (2017--) end-to-end neural systems that jointly learn encoder, quantizer and decoder~\cite{kankanahalli2018,Zhen2019,zhen2020cq,kleijn2021generative,zeghidour2021soundstream}. These new neural audio and speech codecs are often compared to traditional, DSP-based approaches, like open-source  Opus~\cite{valin2012opus}, 3GPP's EVS~\cite{dietz2015evs} and ETSI's LC3/LC3+ codecs. It is worth mentioning that legacy or neural codecs usually code single audio channels. Stereo or multi-channel streams are assembled by aggregating single audio channel instances. Multi-channel codecs, like~\cite{backstrom2021pyawnes}, also appear and this presented proposal is also valid for them.

This paper proposes the Open Audio Coding Evaluation (OpenACE) benchmark, an open-source audio and speech coding evaluation that aims to democratize novel neural or hybrid audio coding and contribute to better evaluation of modern codecs. We design and describe a joint audio and speech coding benchmark that unifies various content types, such as speech and music. We release the benchmark with an open-source evaluation code to further facilitate its use.

We provide two examples of codec comparisons. The first one compares Opus and EVS with two recent LC3 and LC3+ codecs used in Bluetooth Low Energy (BLE) Audio profiles, at varying bitrates using objective intrusive metrics. LC3 was designed for hearing aids and very low power devices. Although BLE Audio gains popularity in consumer electronics, LC3 assessment has been done yet only on small proprietary data. The second example focuses on the emotional speech encoding at 16 kbps, where we show high encoded speech quality dependency on particular emotions.

% The paper is structured as follows. Section~\ref{design} introduces the design principles used to the benchmark data selection. Section~\ref{experiments} describes experiments on how the benchmark could be used, followed by obtained results in Section~\ref{results}. Section~\ref{conclusions} concludes the paper and outlines future work.

%Logi spreadsheet
%https://docs.google.com/spreadsheets/d/1RAFl4Hl4Soailo6XLiys-ELc0ELQ2hAbndYzxJYZriI/edit?pli=1#gid=0

% \footnotesize
\begin{table*}[t!]
\caption{The OpenACE description and statistics. AE: North American English, AS: American Spanish, BE: British English, GR:German, MA:Mandarin, FR:French, DU:Dutch, FI:Finnish, IT:Italian, JP:Japanese, PL:Polish, (B/C/D):ITU-T P.501 Clause, m: male, f: female, a:artificial. Overall, the OpenACE test contains 5.9 hours from 77 speakers with 10 languages.}
\centering
\begin{tabular}{ccccccc}
\toprule
Dataset & Languages & Samp. freq. & Bit & Speakers & Samples & Av. duration \\
& & [kHz] & depth & & & [s] \\
\midrule
IEEE 269-2010~\cite{IEEE269-2010} & AE & 44.1/48 & 16 & 4m, 4f, a & 6 & 7.4 \\
ETSI TS 103-281~\cite{ETSI-TS-103-281} & AE,GR,MA & 48 & 16 & 10m, 10f & 3 & 81 \\
%ETSI TS 103-106~\cite{ETSI-TS-103-106} & AE,GE,MA & 48 & 10m, 10f & 1 & 81\\
%ITU-T P.50~\cite{ITU-T-P50} & AE,BE,FR & 16 & 24m, 24f & 48 & 11\\
ITU-T P.501~\cite{ITU-T-P501} & MA,DU,BE,AE,FR,GR,IT,JP,PL,AS & 8/16/48 & 16 & 20m, 20f & B:48 & B:9\\
& & & & & C:124 & C:8\\
& & & & & D:30 & D:6\\
%find . -name "*.wav" | xargs -I@ bash -c "soxi -D '@'"
VCTK test~\cite{valentini2017noisy} & English with various accents & 48 & 16 & 3m & 831 & 2.5\\
% Common Voice~\cite{CommVoice7} & English with various accents & 48 & - & 2290 & -\\
%cat en/test.tsv | cut -f6 | grep -v -e '^$' | wc -l
%      2 eighties
%    110 fifties
%    215 fourties
%     33 seventies
%     46 sixties
%    318 teens
%    445 thirties
%   1121 twenties
% MagnaTagATune~\cite{law2009evaluation} & - & - & - & - & -\\
% FSDKaggle2018~\cite{fonseca2018general} & - & 44.1 & - & XX & XX\\
EARS test ~\cite{richter2024ears} & English with various accents & 48 & 32 & 3m, 3f & 886 & 15 \\
\midrule
EBU SQAM~\cite{EBU-SQAM} & - (Audio) & 44.1 & 16 & - & 70 & 50 \\
ODAQ~\cite{Torcoli_icassp24} & - (Audio) & 44.1 & 24 & - & 47 & 11 \\
\bottomrule
\end{tabular}
\label{tab:benchmarkdata}
\end{table*}
\normalsize

\section{Benchmark proposal}
\label{design}
\subsection{Design Principles}

% \begin{itemize}
%     \item Sampling frequencies
%     \item Data availability
%     \item Content type, etc
% \end{itemize}

\textbf{Sampling frequency and bit depth:} 
Modern codecs can scale from narrowband (8kHz) speech to high-quality full-band (48kHz) stereo audio, with bits per audio sample ranging from 16 to 32 bits. Internal calculations in LC3 is 16 bits, and 24 bits in LC3+. Most open speech data is sampled at 16kHz, traditionally enough for automatic speech recognition. However, the proposed coding benchmark should thus contain full-band audio and speech data with higher bit depths that allow downsampling to the target evaluation setting. 
% %Clean speech, noise and room impulse responses sampled at both required rates are available from the Interspeech 2021 Deep Noise Suppression Challenge~\cite{reddy21interspeech}.

\textbf{Legacy and availability:} 3GPP, IEFT or ETSI codecs are extensively tested but with restricted access data. Even the open-source Opus project provides the testing vectors\footnote{\url{https://opus-codec.org/testvectors}} to verify the codec implementation but does not provide the evaluation data. Any benchmark proposal should opt for open-source data to democratize the evaluation and reproducibility of the candidate systems. There are some available test signals from the standardization organizations that should be included in any speech coding testing. Table~\ref{tab:benchmarkdata} lists the three such datasets from the top.

% %The IEEE 269-2010~\cite{IEEE269-2010} contains six original recordings that were made with a 1/2 measurement microphone in a quiet and absorbent room. The signals were originally recorded with 48 kHz sampling rate. The ETSI TS 103 281~\cite{ETSI-TS-103-281} contains the American English, German and Mandarin test signals.

\textbf{Data variability:} As mentioned in the introduction, speech and audio coding are used ubiquitously worldwide and for many media types. Candidate codecs need to demonstrate robustness to i) intrinsic signal variabilities that include speakers: \textit{language}, \textit{age}, media type: \textit{typical and atypical speech}, \textit{audio events} and \textit{music}, and ii) extrinsic signal variabilities including environment: \textit{noisy} and \textit{reverberant}, and transmission errors caused, for example, by the \textit{channel} and \textit{RF interference}.

% BLE use cases: voice, music streaming, hearing aid (WB or semi-SWB, 16 or 24 kHz) and broadcast audio (WB, S-SWB, FB).

% BLE codec attributes: i) bitrates: from 16 kbps to 320 kbps (20 bytes to 400 bytes per 10-ms frames), ii) complexity (FLOPS?) - skip in this paper, iii) System Delay, iv) multi-channel coding, and v) Packet loss concealment.

% BLE listening tests: expert or non-expert listeners?, with or without the transmission error?, and three methods were used: i) P.800 Absolute Category Rating (ACR) - for packet loss, ii) MUlti Stimulus test with Hidden Reference and Anchor (MUSHRA) and ITU-R BS.1116-3. The material is from the European Broadcast Union (EBU) test material Table 2.2, Table 2.4, Table 2.5, Table 2.7 and 2.8 from European Broadcast Union (EBU), “EBU – TECH 3253 - Sound Quality Assessment Material recordings for subjective tests", September 2008, \url{https://tech.ebu.ch/docs/tech/tech3253.pdf} - data available from \url{https://tech.ebu.ch/publications/sqamcd} - UK (m/f), French (m/f) and German (m/f), 6 spks in total, each recording 22 seconds long.

% Shuld the audio benchmark data  include: near-field (boom and boomless microphones), far field, with and without speech enhancement?

\textbf{Speech quality assessment:} The ITU-T standards Perceptual Evaluation of Speech Quality (PESQ), Perceptual Objective Listening Quality Analysis (POLQA), and ETSI's 3-fold Quality Evaluation of Speech in Telecommunications (3QUEST) belong to the most popular intrusive speech quality methods. Clean references are required to compute the objective scores. To perform subjective/listening test correctly, the speech files need to be calibrated to an active speech level of 79 dB SPL. 

% %\textbf{Speech intelligibility assessment:} Short-Time Objective Intelligibility (STOI).

% \textbf{Audio Quality Assessment:} Comparison of different audio quality assessment methods~\cite{zacharov2017comparison}

% \textbf{Augmentation types:} Various distortion should be available for simulate distortions. The degraded speech samples can be obtained by an application of various Voice Activity Detectors (VADs), noise supression algorithms, network/packet loss scenarios and handset/hands-free modes.

% \textbf{Real-world data:} The intrusive quality methods can be used but with many limitations. For example, POLQA processor requires 
% at least ~3 seconds of speech. Ideally, the testing speech segments should be 8-12 seconds long. In each segment there should be 2-3 bursts of speech of about 3-4 seconds each, and each burst should be separated by pauses of 1-2 seconds. Besides, POLQA was never intended for evaluating reverberant speech. According to ITU-T Recommendation P.863.1~\cite{p863.1}, T60 should be below 0.3 seconds above 200Hz, with the distance to the microphone of approximately 10cm. Real-world data often does not fit such the limitations, and non-intrusive methods have to be used. The JAQOB benchmark thus also includes the real-world data.

\subsection{Benchmark data}

In this work, we propose a joint audio-speech content types quality benchmark, mixing traditional open assessment datasets with recent large, higher quality, and more variable content datasets:
\begin{itemize}
    \item The IEEE 269-2010~\cite{IEEE269-2010} standardizes methods for making laboratory measurements of electroacoustic characteristics of analog and digital telephones, handsets, and headsets. The methods may also apply to various other communications equipment, including cordless, wireless, and mobile communications devices. It includes a small available full-band dataset of 8 speakers.
    \item The ETSI TS 103-281~\cite{ETSI-TS-103-281} provides the test audio vectors from 20 speakers in 3 languages, designed to be used to objectively evaluate the performance of super-wideband and full band mobile terminals for speech communication in the presence of background noise.
    \item  The ITU-T P.501~\cite{ITU-T-P501} recommendation gives a wide variety of test signals, starting with low-complexity test signals up to test signals with a high degree of complexity incorporating many typical parameters of speech. It contains multi-lingual test samples from ten different languages and sampling frequency variation from 8 to 48 kHz.% On contrary, only 16 bits are used for samples encoding.
    \item The VCTK test samples~\cite{valentini2017noisy} are widely used in neural or hybrid speech coding proposals, as in 2017, not many full-band high-quality data were opened for researchers. Initially designed for voice cloning experiments, it includes many utterances from fewer speakers.
    \item The EARS, anechoic fullband speech dataset~\cite{richter2024ears}, was originally benchmarked for speech enhancement and dereverberation. It is also excellent for speech coding benchmarking, as it includes a large range of different speaking styles, including emotional speech, reading styles, non-verbal sounds, and conversational freeform speech. We included all six test speakers.
    \item Two audio datasets: EBU-SQAM~\cite{EBU-SQAM} and ODAQ~\cite{Torcoli_icassp24}. The latter one includes the audio material from newly released productions by Netflix and Fraunhofer IIS at 24-bit sample resolution.
\end{itemize}
% The ITU-T P.501~\cite{ITU-T-P501} contains the multi-lingual test samples. The VCTK test~\cite{valentini2017noisy} set consists of clean and noisy reverberated speech samples. Common Voice 7.0~\cite{CommVoice7} test set contains the age labels that varies from teens to eighties.

% \begin{itemize}
%     \item Introduction and description of Tab.~\ref{tab:benchmarkdata}.
% \end{itemize}

% Notes
% Codecs: 
% Franhofer ACC (250kbps)
% Sony LDAC (990 kpbs)
% Qualcomm aptX (352 and 576 kpbs)
% SBC (sub-band codec) 300 kbps
% A2DP advance audio distribution profile
% low fidelity codecs (~16 kHz)
% LC3 codec (BLE audio) - Fraunhofer contributed 160 kbps+, multi-channel support, 5ms latency

% ITU-T Perceptual Speech Quality ITU-T P.835

% https://opus-codec.org/testvectors/

% running Opus and EVS:
% https://github.com/xiph/opus
% https://github.com/vipchengrui/EVS-codec/tree/master/source_code

% https://www.iis.fraunhofer.de/en/ff/amm/communication/lc3.html
% https://www.etsi.org/deliver/etsi_ts/103600_103699/103634/01.02.01_60/ts_103634v010201p0.zip

% sudo apt-get install opus-tools
% opusenc, opusdec speech 48kHz 6-21 kbps

\begin{table*}
\caption{The VISQOL and POLQA scores computed for reference and encoded audio on the full-band subset of the benchmark data.
% POLQA results are shown for a selected subset of the full band data because of computational constraints.
The input files were truncated to the first 10 seconds. 95\% confidence intervals are computed using 1000 bootstrap samples.}
\centering
\begin{tabular}{ccc|cc|cc}
\toprule
     & \multicolumn{2}{c}{16 kbps} & \multicolumn{2}{c}{32 kbps} & \multicolumn{2}{c}{64 kbps} \\ \midrule
     & VISQOL   & POLQA    & VISQOL   & POLQA& VISQOL   & POLQA (subset)   \\ \midrule
LC3  & 3.51 [3.49, 3.52]     & 2.98 [2.95, 3.01]
& 4.00 [3.99, 4.01]     & 4.44 [4.42, 4.46]& \textbf{4.39 [4.38, 4.40]}     & 4.64 [4.63, 4.65]\\
LC3+ & 3.11 [3.08, 3.13]     & 3.23 [3.20, 3.26]
& 4.00 [3.99, 4.01]     & 4.45 [4.43, 4.47]& \textbf{4.39 [4.38, 4.40]}     & \textbf{4.64 [4.63, 4.65]}
\\
EVS  & 3.92 [3.90, 3.93]     & \textbf{4.34 [4.32, 4.35]} & 3.82 [3.81, 3.84]     & 4.37 [4.36, 4.39]& 4.19 [4.17, 4.20]     & 4.53 [4.52, 4.55]
\\
Opus & \textbf{3.98 [3.97, 3.99]}     & 4.03 [4.01, 4.05]
& \textbf{4.05 [4.04, 4.06]}     & \textbf{4.56 [4.54, 4.58]}& 4.37 [4.36, 4.38]     & 4.57 [4.56, 4.58]\\ \bottomrule
\end{tabular}
\label{tab:objective_results}
\end{table*}

\section{Experiments}
\label{experiments}
% Compare SBC, G.722, Opus, Lyra, EnCodec, LC3 and LC3 plus:
% \begin{itemize}
%     \item LC3: \url{(https://github.com/google/liblc3}
%     \item LC3+: \url{https://www.etsi.org/deliver/etsi_ts/103600_103699/103634/01.03.01_60/ts_103634v010301p0.zip}
%     \item Opus: \url{https://github.com/xiph/opus}
%     \item EVS: \url{https://github.com/vipchengrui/EVS-codec/tree/master/source_code}
%     \item \url{https://www.iis.fraunhofer.de/en/ff/amm/communication/lc3.html}
%     \item LMCodec (ICASSP23): \url{https://arxiv.org/pdf/2303.12984.pdf}
%     \item Lyra (Google SoundStream): \url{https://github.com/google/lyra}
%     \item EnCodec (Meta): \url{https://github.com/facebookresearch/encodec}
%     \item free ultra-band 96khz test music tracks \url{https://www.soundonsound.com/techniques/sos-audio-test-files-downloads}
% \end{itemize}

This section describes the experimental setup of our two-example usage of the OpenACE benchmark. We focus on two particular use cases: objective evaluation of audio quality coding performance leveraging intrusive metrics and subjective evaluation through listening tests. In particular, we compare four DSP-based codecs: Opus \cite{opustools}, EVS \cite{EVS_imp}, LC3 \cite{liblc3}, and LC3+ \cite{lc3_plus_imp}.
% LC3 supports bitrates from 16 kbps to 320 kbps for 10-ms frames.
Existing evaluation~\cite{lc3white} reports LC3 performance on a test set of 24 German language samples only and misses assessment of the minimal supported bitrate of 16 kbps. Using the OpenACE, the size of the test increases by several orders of magnitude. The experiments only evaluated channel conditions without transmission errors, such as packet losses. The used signals are prepared by downmixing them to mono and converting the audio to a unified 16-bit PCM WAV format.

\subsection{Objective evaluation of audio quality vs bitrates}

We leverage the intrusive VISQOL Audio \cite{Chinen2020ViSQOLVA} and POLQA \cite{p863.1} metrics for the objective evaluation. NISQA~\cite{mittag21_interspeech} that supports full-band signals is also available in the project repository. Given the 48 kHz sample rate requirement of the VISQOL algorithm, we select the full-band signals in the benchmark and resample them from 44.1 to 48 kHz using Kaiser window resampling. For objective metric computation, we truncate the audio to the first 10 seconds of each file. We evaluate the four coding algorithms at 16, 32, and 64 kbps. 

Table \ref{tab:objective_results} presents the objective scores for the various coding methods at the three different bitrates. The results show that at the low bitrate setting of 16 kbps, LC3+ shows the lowest performance, with LC3 performing slightly better. Notably, there is a significant gap of 0.4 between the two LC3-based codecs. EVS and Opus, on the other hand, achieve similar, higher results, with Opus having the highest score. Our results complement the previous LC3 assessment~\cite{lc3white} that did not report the results obtained at 16 kbps. It is also fair to add that LC3 complexity encoding/decoding is significantly lower than Opus complexity. For example, in Hearing Aid music task, Opus has about twice as much complexity. %This, however, results in degraded LC3 quality in lower bitrates.

For 32 kbps, we see an improvement in scores for Opus, LC3, and LC3+. Large quality increases were obtained for the LC3-based codecs, with LC3 and LC3+ gaining 0.49 and 0.89, respectively. On the other hand, Opus only obtained a slight increase of 0.07 but retained the best performance at 32 kbps. Surprisingly, the objective scores for EVS decreased slightly by increasing the bitrate from 16 to 32 kbps.  

The results for the highest bitrate of 64 kbps show a significant increase in objective scores for all codecs. All methods obtain an approximate increase of 0.3 over the 32 kbps condition. At this bitrate, we can also observe that the two LC3-based codecs overtook EVS and Opus and obtained the highest performance of 4.39 MOS.  

\subsection{Subjective evaluation of emotional speech at 16 kbps}

The second experiment, conducted through subjective assessment, focused on studying the coding performance of emotional speech. To this end, we selected from the OpenACE a subset of five basic emotions: sadness, ecstasy, pain, fear, and anger. For reference, we added neutral (non-emotional) speech. The selected recordings were from the freeform section of the dataset. All of the tested codecs were set to operate at 16 kbps.

To evaluate the subjective quality of the different coding methods, we conduct an ITU-R BS.1534 MUSHRA test \cite{BS_1534_3}. We used low- (3500 Hz) and mid-quality (7000 Hz) anchors generated using a type-1 Chebyshev filter. We truncated listening stimuli to the first 6 seconds.

\begin{table}[ht]
\caption{Mean MUSHRA scores of emotional speech encoding quality with 95\% confidence intervals, obtained across all codecs.}
\centering
% \resizebox{\linewidth}{!}{
\begin{tabular}{c|c}
\toprule
Emotion & Rating \\
\midrule
Neutral & 63.61 [61.26, 65.96]\\
\midrule
Sadness & 64.27 [61.95, 66.59]\\
Fear & 63.14 [60.77, 65.51]\\
Pain & 61.75 [59.32, 64.18]\\
Ecstasy & 60.90 [58.50, 63.30]\\
Anger & 60.51 [58.11, 62.91]\\
\bottomrule
\end{tabular}
% }
\label{tab:emo}
\end{table}
\normalsize

% \begin{figure}[ht]
% \centering
% \includegraphics[width=\linewidth]{Fig1.png}
% \caption{Overall, neutral and sad speech were better encoded with 64.1 $\pm$ 2.6 ratings, slightly above fear, which resulted in 63.8 $\pm$ 2.7. Pain (61.4 $\pm$ 2.7) with Anger (60.8 $\pm$ 2.7) speech were more challenging, mostly for LC3 and LC3+ codecs.}
% \label{fig:emo1}
% \end{figure}

Using the headphones, 36 listeners conducted the MUSHRA test through an online platform\footnote{\url{https://senselabonline.com/}}. Tab.~\ref{tab:emo} shows obtained quality aggregated over all tested codecs. Type III ANOVA test revealed that there is a statistically significant difference in emotion encoding quality (F value = 8, $p<0.001$). We thus confirm that audio coding degrades signal quality, as reported by the EMO-Codec benchmark~\cite{ren2024emo}. Their authors found that both DSP and ML-based codecs do not encode emotional speech sufficiently.
% , which is more challenging in lower (below 24 kbps) bitrates.
Besides, the used IEMOCAP dataset~\cite{busso2008iemocap} needs to be licensed, which is not required with the OpenACE data. The test data is available here\footnote{\url{https://huggingface.co/datasets/mcernak/EARS-EMO-OpenACE}}.

\begin{figure*}[ht]
\centering
\includegraphics[width=\linewidth]{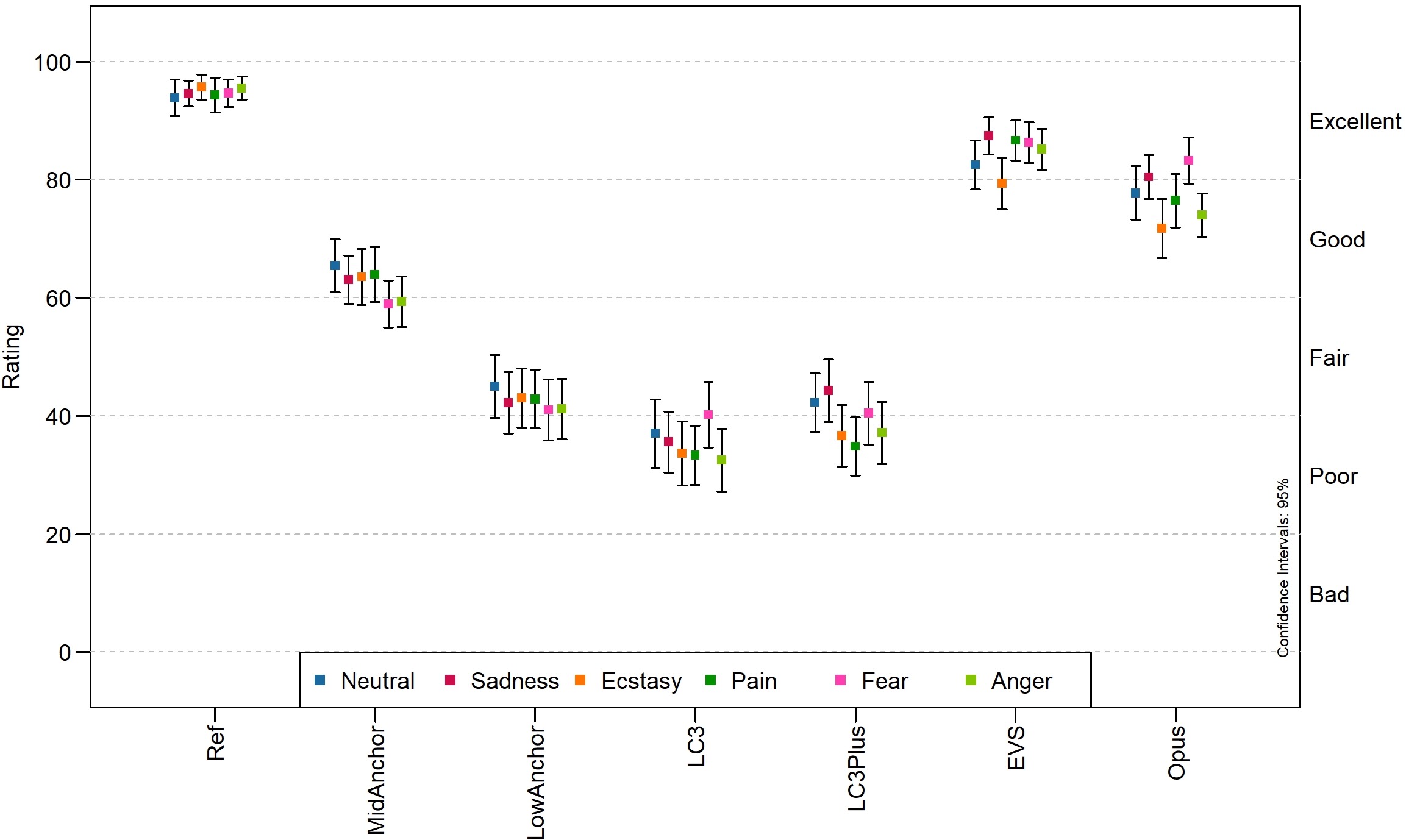}
\caption{Mean overall speech quality with 95\% confidence intervals obtained of emotional speech encoding using the MUSHRA test~\cite{BS_1534_3}.}
\label{fig:emo2}
\end{figure*}

Fig.~\ref{fig:emo2} shows mean overall speech quality split across the encoding methods and emotional speech categories. The subjective test confirms the overall lower performances of LC3 and LC3+ codecs on the lower bitrate of 16 kbps. EVS usually provides better quality at the same bitrate as Opus~\cite{evs-eval2015}, and we confirmed it also for emotional speech.

\section{Discussion}
\label{discussion}
The results for the objective assessment of the Opus, EVS, LC3, and LC3+ codecs on full-band audio highlight the differences between the codecs for varying bitrates. We observed that the LC3-based codecs performed significantly worse at lower bitrates than Opus and EVS. The results at 16 kbps also show that LC3+ performs worse than the standard LC3. This difference in measured performance is likely due to the difference in default behavior between the codecs, as the LC3+ codec applies a low-pass filter at 12 kHz, while LC3 applies a 20 kHz low-pass filter. Comparing the objective and subjective test results for LC3 and LC3+ reveals a discrepancy between the predicted objective scores and human evaluations. At 16 kbps, the objective results show a significant performance gap between LC3 and LC3+, which can not observed in the subjective test results, even when computing the VISQOL scores on only the subset used in the MUSRA test. This suggests that VISQOL may overemphasize the impact of lowpass filtering over human perception.

Audio coding has seen a paradigm shift to data-driven approaches, and neural encoding/decoding used in multimodal large language models will emphasize this research direction even more in the future. On the contrary, delay and complexity are intrinsic requirements for on-device processing, and further work will be done on ML-based codec models' compression. All those developments can benefit from common benchmarking, such as the proposed OpenACE.

\section{Conclusions}
\label{conclusions}
We have described the audio and speech coding data that consist of 3 legacy coding data from the IEEE, ETSI and ITU-T, two open popular test data (VCTK and EBU SQAM), and two recent high-quality data released in 2024 (EARS and ODAQ). We have shown four DSP-based codecs benchmarking at 16 -- 64 kbps and demonstrated their performance on emotional speech audio. The implemented open-source DSP-based codec evaluation can be easily used as the baseline method for current and future neural or hybrid codec R\&D.

In the future, we want to maintain and extend the benchmark. Namely, we aim to add data augmentation like environmental noises and reverberation, and physical channel simulations for transmission error modeling, such as radio frequency interference. Such channel data augmentations may contribute to end-to-end quality/delay/complexity codec assessment. Adding multi-channel array(s) modeling will approach the codec benchmark to real-life conditions.

\newpage
\clearpage
\newpage

\balance 

\footnotesize
\bibliographystyle{IEEEbib}
\bibliography{references}

\end{document}